\documentclass[12pt,preprint]{aastex}
\shorttitle{CV in M22}
\shortauthors{Bond et al.}

\begin{document}


\title{Multiple Outbursts of a Cataclysmic Variable in the Globular Cluster M22}


\author{
I.A. Bond\altaffilmark{1}, 
F. Abe\altaffilmark{2}, 
S. Eguchi\altaffilmark{2}, 
Y. Furuta\altaffilmark{2}, 
J.B. Hearnshaw\altaffilmark{3}, 
K. Kamiya\altaffilmark{2}, 
P.M. Kilmartin\altaffilmark{3}, 
Y. Kurata\altaffilmark{2}, 
K. Masuda\altaffilmark{2}, 
Y. Matsubara\altaffilmark{2}, 
Y. Muraki\altaffilmark{2}, 
S. Noda\altaffilmark{4}, 
K. Okajima\altaffilmark{2}, 
N.J. Rattenbury\altaffilmark{5}, 
T. Sako\altaffilmark{2}, 
T. Sekiguchi\altaffilmark{2},  
D.J. Sullivan\altaffilmark{6}, 
T. Sumi\altaffilmark{7}, 
P.J. Tristram\altaffilmark{9},
T. Yanagisawa\altaffilmark{8}, 
and P.C.M. Yock\altaffilmark{9}
\\(The MOA Collaboration)
}

\altaffiltext{1}{Institute for Information and Mathematical Sciences,
Massey University, Auckland, New Zealand; i.a.bond@massey.ac.nz}
\altaffiltext{2}{Solar-Terrestrial Environment Laboratory, Nagoya University, Nagoya 464-8601, Japan;
\{abe, sada, furuta, kkamiya, kurata, kmasuda, ymatsu, muraki, sako, sekiguchi\}@stelab.nagoya-u.ac.jp}
\altaffiltext{3}{Department of Physics and Astronomy, University of Canterbury, Private Bag 4800, 
Christchurch, New Zealand; \{john.hearnshaw, pam.kilmartin\}@canterbury.ac.nz}
\altaffiltext{4}{National Astronomical Observatory of Japan, Tokyo, Japan; sachi.t.noda@nao.ac.jp}
\altaffiltext{5}{Jodrell Bank Observatory, Macclesfield, Cheshire SK11 9DL, UK; njr@jb.man.ac.uk}
\altaffiltext{6}{School of Chemical and Physical Sciences, Victoria University of Wellington, PO Box 600, Wellington, 
New Zealand; denis.sullivan@vuw.ac.nz}
\altaffiltext{7}{Department of Astrophysical Sciences, Princeton University, Princeton NJ 08544, USA;
sumi@astro.princeton.edu}
\altaffiltext{8}{National Aerospace Laboratory, Tokyo 182-8522, Japan; tyanagi@nal.go.jp}
\altaffiltext{9}{Department of Physics, University of Auckland, Auckland, New Zealand;
p.yock@auckland.ac.nz, pjt61@ext.canterbury.ac.nz}


\begin{abstract}
We present a 4 year light curve of a cataclysmic variable in M22, based on an 
analysis of accumulated data from the MOA microlensing survey. The position of the
star coincides with that of a transient event observed by HST in 1999, originally attributed
to microlensing but later suspected to be a dwarf nova outburst.
Two outburst episodes, one in 2002 and one in 2003,
with $\Delta I\sim3$ are seen in the MOA data thus confirming that the HST 
event was a dwarf nova outburst. The MOA and HST data show that this dwarf nova underwent 
at least three outburst episodes during 1999--2004. Further close monitoring of this 
event is encouraged as future outburst episodes are expected.
\end{abstract}


%% We move on to the body of the paper.

\section{Introduction}

Eruptions of cataclysmic variables (CVs) and microlensing are two types of 
astronomical transient phenomena that are of considerable interest if they are 
associated with globular clusters. Through their identification with compact binaries, CVs play
an important role in the dynamical evolution due to their large binding energy relative to that
of the cluster as a whole \citep{hut92}. The frequency of outbursts of CVs may play an additional
role through the energy releases involved. However, while dozens of CVs in quiescence have now been identified 
in globular clusters \citep{pooley02}, very few have been observed in outburst. On the other hand, in 
microlensing events, globular clusters can provide source stars for foreground lenses
or lenses for background source stars.
Such events provide probes of dark matter either as Massive Compact Halo
Objects (MACHOS) or as dark objects bound within the clusters themselves \citep{jetzer98}.

In 1999, observations of M22 were carried out
by the HST in a program designed to search for microlensing events towards this
globular cluster. In an analysis of these data, \citet{sahu01} reported one event characterized by a
brightening and fading of a star by about 3 magnitudes over a 20 day timescale.
Having noted that the light curve of the star at baseline featured a $\sim$30 day
modulation of amplitude 0.1 mag, they interpreted this event to be microlensing of
a variable field star in the Galactic Bulge by an unseen foreground lensing object in the
cluster. However, in an examination of additional archival HST data, \citet{anderson03} argued that the
photometry in the $V$, $I$, $R$, and $H\alpha$ passbands, together with the derived 
X-ray luminosity measured in ROSAT/HRI images, all point to the star in question
having the same optical and X-ray signatures of a CV in quiescence.  They also found that the proper motion of the star 
matched the mean proper motion of other stars in the cluster and thus argued that the star 
associated with the event is also in M22. They then inferred
that the episode reported by \citet{sahu01} was probably an outburst of a dwarf nova
cataclysmic variable located in M22.

%% Why important to sort this out

Since 2000, the MOA (Microlensing Observations in Astrophysics) collaboration has been
regularly monitoring a number of fields in the Galactic Bulge as part of its microlensing
survey program. In mid-2001, a new survey field centred on M22 was added to the program.
The accumulated data set therefore provides a useful opportunity to check for possible repeat
outbursts and to settle the matter regarding its nature. In this letter
we present an analysis of this dataset and extract a 3.5 year light curve where we indeed see 
two outburst episodes. 

\section{Observations and Data}

The MOA program is carried out using a 60 cm telescope at the Mt John Observatory in New 
Zealand. Observations of fields towards the Galactic Bulge, including M22, run from February
to November in each year. The camera comprises three 2K$\times$4K CCD chips giving a total field
of view of (1.4\degr$\times$0.9\degr). The observation images are mainly 180 s exposures through a 
broadband $I$ filter. The images are reduced in real-time
using difference imaging to optimise the effectiveness of detecting microlensing and other 
transient events \citep{bond01}. In this process, no transient events were detected near the 
position of the reported outburst event in M22. This is not unexpected as this star is located in 
a very dense region crowded with saturated stars and any outburst could have easily been missed
by the real-time detection system. We therefore decided to carry out a careful follow-up analysis
of "cameo" images derived by extracting 801$\times$801 pixel sub-rasters (corresponding to a 
field-of-view of $\approx$11\arcmin), centred approximately 
on the event. Observations where the seeing was 
worse than 2.7$\arcsec$ were rejected since the effects of saturated
stars would wash out any signal from the star of interest. This left us with 440 good quality
cameo images spanning the period from June 2001 to November 2004. 
All observation images were 
treated with the usual bias subtraction and flat-fielding techniques before extracting 
cameo images. 

\section{Analysis and Results}

The analysis of these data takes the practice of crowded field photometry to its extreme
limit. We do not expect to be able to resolve the source star at baseline, but with a 
careful application of difference imaging photometry, we can expect to be able to detect
the star during any suitably bright outburst episodes. Our difference imaging procedure,
described in detail by \citet{bond01}, is as follows. Given a set of time series images of a 
particular field in the sky, one first chooses a "reference" image amongst those with the best seeing and 
signal to noise ratio. The reference is then subtracted from each of the remaining images
after applying a convolution operation to the reference image to match the seeing of the 
other image. The convolution operator is modelled directly using the techniques of \citet{al} and 
\citet{alard}. Our set of cameo images was put this pipeline to produce a set of 439
difference images.

\begin{figure}
%\epsscale{.30}
\plotone{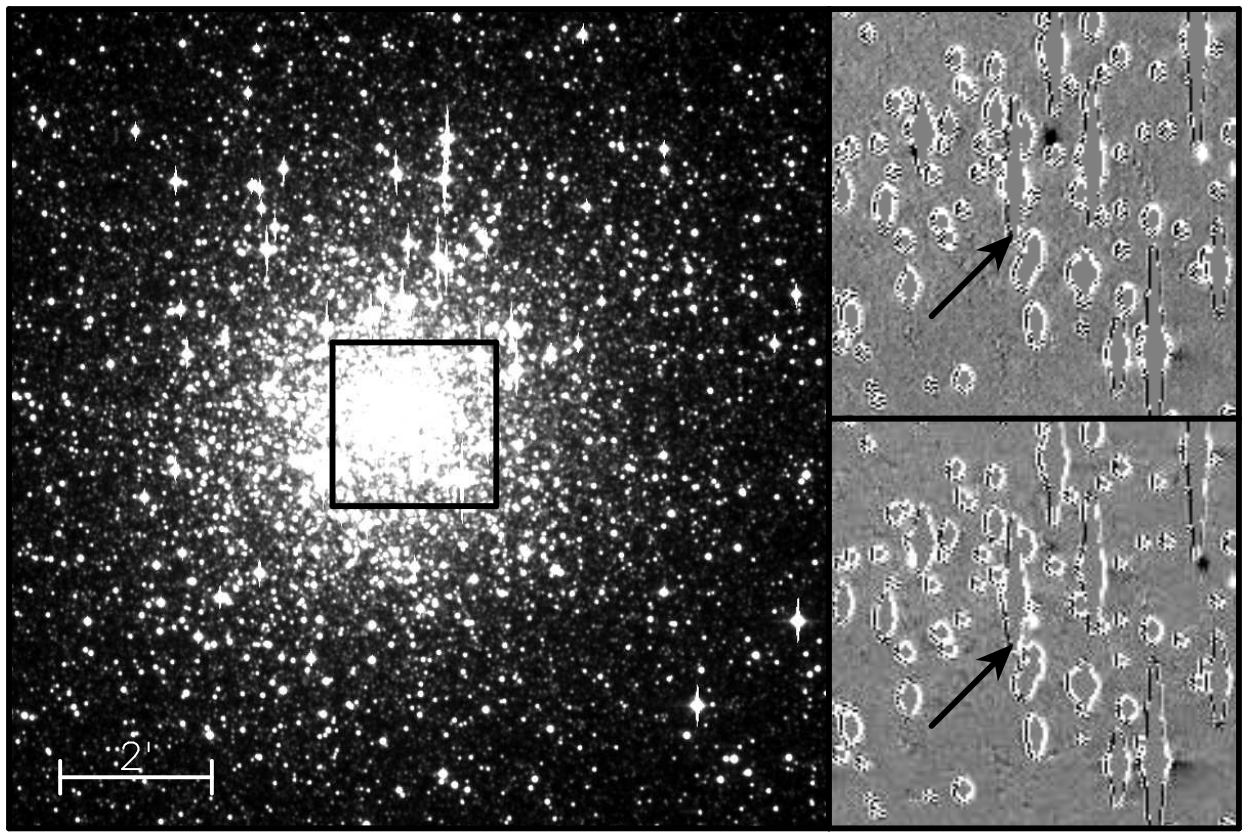}
\caption{MOA images of M22. The large image shows the reference image selected from the set of cameo images (refer text) 
extracted from MOA observations of M22. The smaller images show two difference images corresponding to the subsection
on the reference image. The spoiling effects of saturated stars can be seen. The arrow in each difference image
point to the position of the event reported by \citet{sahu01}. An outburst epsiode can be seen in the lower image.\label{fig1}}
\end{figure}

In Fig.~\ref{fig1}, we show the reference image together with two difference images---one of which
shows a stellar-like image at the position reported by \citet{sahu01}. In difference imaging 
analysis, these profiles indicate variability associated with a resolved or unresolved star. 
Although the object indicated in Fig.~\ref{fig1} is partly affected by the nearby saturated star,
we are confident that this feature is not some artifact that was somehow introduced into the
analysis. The shape of its profile matches that of similar profiles elsewhere on the difference
images that are more clearly identified as being associated with stellar variability. We thus 
identify the feature shown in Fig.~\ref{fig1}, as a brightening episode associted with a star that
is unresolved in the individual observation images. 

\begin{figure}
\plotone{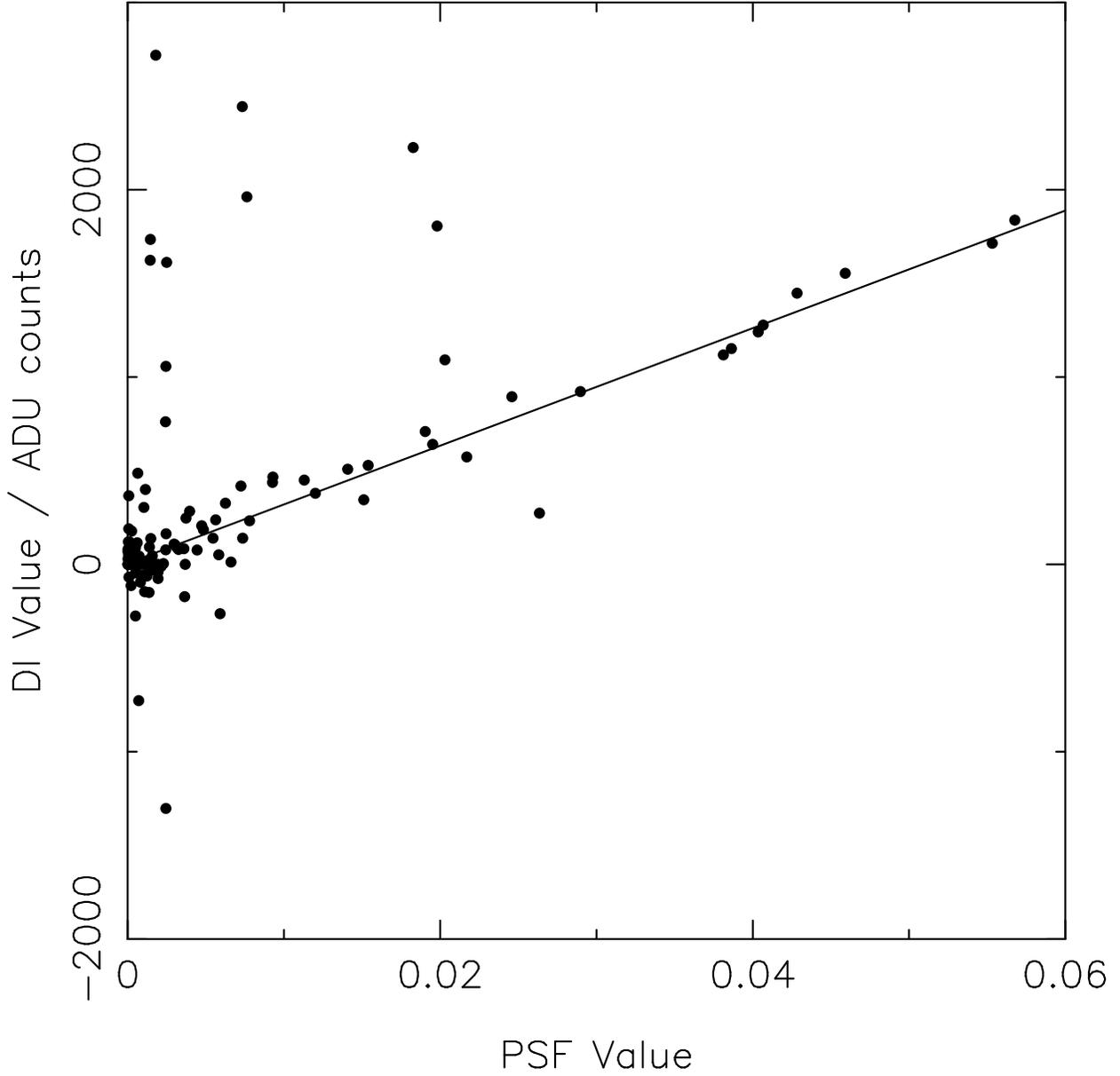}
\caption{PSF fitting to difference images as a pixel by pixel cross plot of a 
normalised PSF located at the position of the source. The signal can be clearly identified 
against the outliers due to nearby saturated pixels. The flux may be extracted using a robust 
linear fit that rejects these outliers, as shown.\label{fig2}}
\end{figure}

\begin{figure}
\plotone{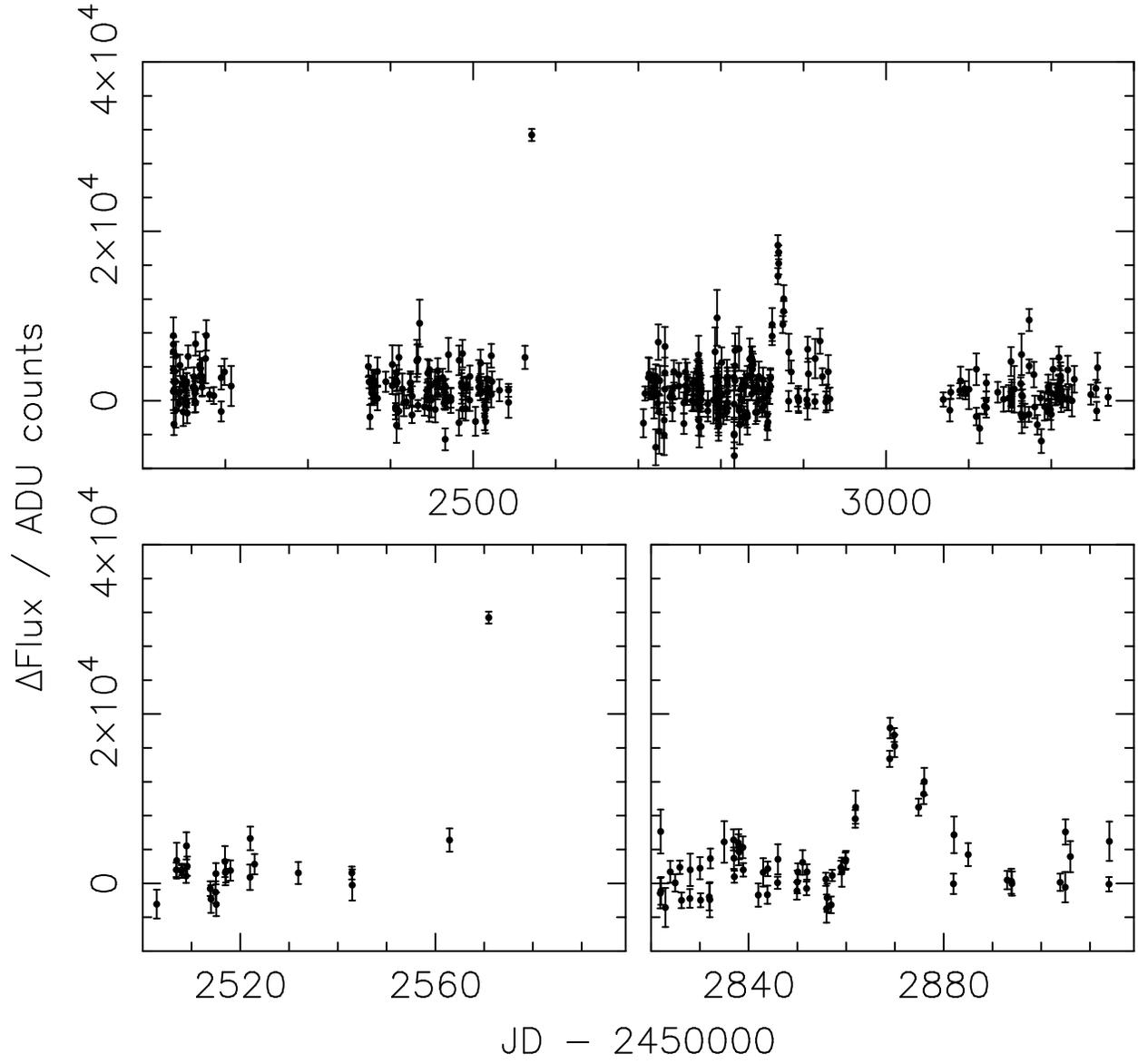}
\caption{Difference imaging light curve during 2001---2004 of the event in M22 shown in Fig.~\ref{fig1}. 
Two episodes of brightening, shown in the lower panels, were identified on the difference 
images. \label{fig3}}
\end{figure}

We measured the centroid position of the object shown in Fig.~\ref{fig1}, by fitting a Gaussian profile 
to a set of one dimensional profiles in the two axes. We used this measurement to fix the position of the
object in the coordinate frame of the reference image. To extract flux measurements from all difference images
we first constructed a reference empirical PSF from isolated stars in the reference image. For a given difference
image we then placed the empirical PSF, convolved with the convolution operator for that image, at the object
centroid. The flux is then the slope of a linear fit to the pixel-to-pixel cross plot of the PSF image and 
difference image. The contaminating effects of the nearby saturated stars made it necessary to use robust linear
fitting techniques that remove outliers as shown in the example in Fig.~\ref{fig2}.
The resulting light curve is shown in Fig.~\ref{fig3}. The outburst shown in Fig.~\ref{fig1} occurred at JD2452570 
(2002 Oct 23). This was right at the end of the Galactic Bulge viewing season for 2002 and as such, only 
one sampling point in the outburst profile was obtained.
A second outburst at the event position was detected during JD2452850--2890 (2003 Jul 29--Sep 7). 
Although the amplitude was not as great as that in the first outburst, these measurements were time resolved.

The light curve is presented as "delta flux" measurements with respect to
the flux of the star on the reference image. The quiescent or baseline flux of the star is well
below the MOA detection threshold, so the measurements shown in Fig~\ref{fig3} are effectively
total fluxes. Outside the two outburst epeisodes, the delta flux measurements are scattered 
around the zero point, indicating that the reference image samples the event during baseline 
or quiescence, or another outburst event with an amplitude below the detection threshold. It 
is important to note that outside the outbursts, the measurements shown in Fig.~\ref{fig3} are 
essentially background. Also the light curve only samples the outbursts during those periods 
when the outburst luminosity is above the MOA threshold. It is also possible
that other smaller amplitude outbursts, peaking below the threshold, occurred 
during 2001--2004. 

The flux measurements shown in Fig.~\ref{fig3} are in integrated ADU counts, all photometrically 
scaled to stellar fluxes on the reference image. To determine standardized magnitudes, we applied
the calibration procedure described in \citet{bond01} using the Johnson-Cousins BVI observations 
of \citet{monaco04} as calibration data. In Table~\ref{tab1}, 
we present the corresponding magnitudes from measurements during the two outbursts, where the signal 
to noise was better than 4$\sigma$. From the scatter in the ADU counts in the lightcurve outside the 
outburst events, and our calibration information, we estimate the limiting magnitude
at the position of this event on the image to be $I\sim$17.5 (observations within less dense 
regions of the field would go deeper).

\begin{figure}
\plotone{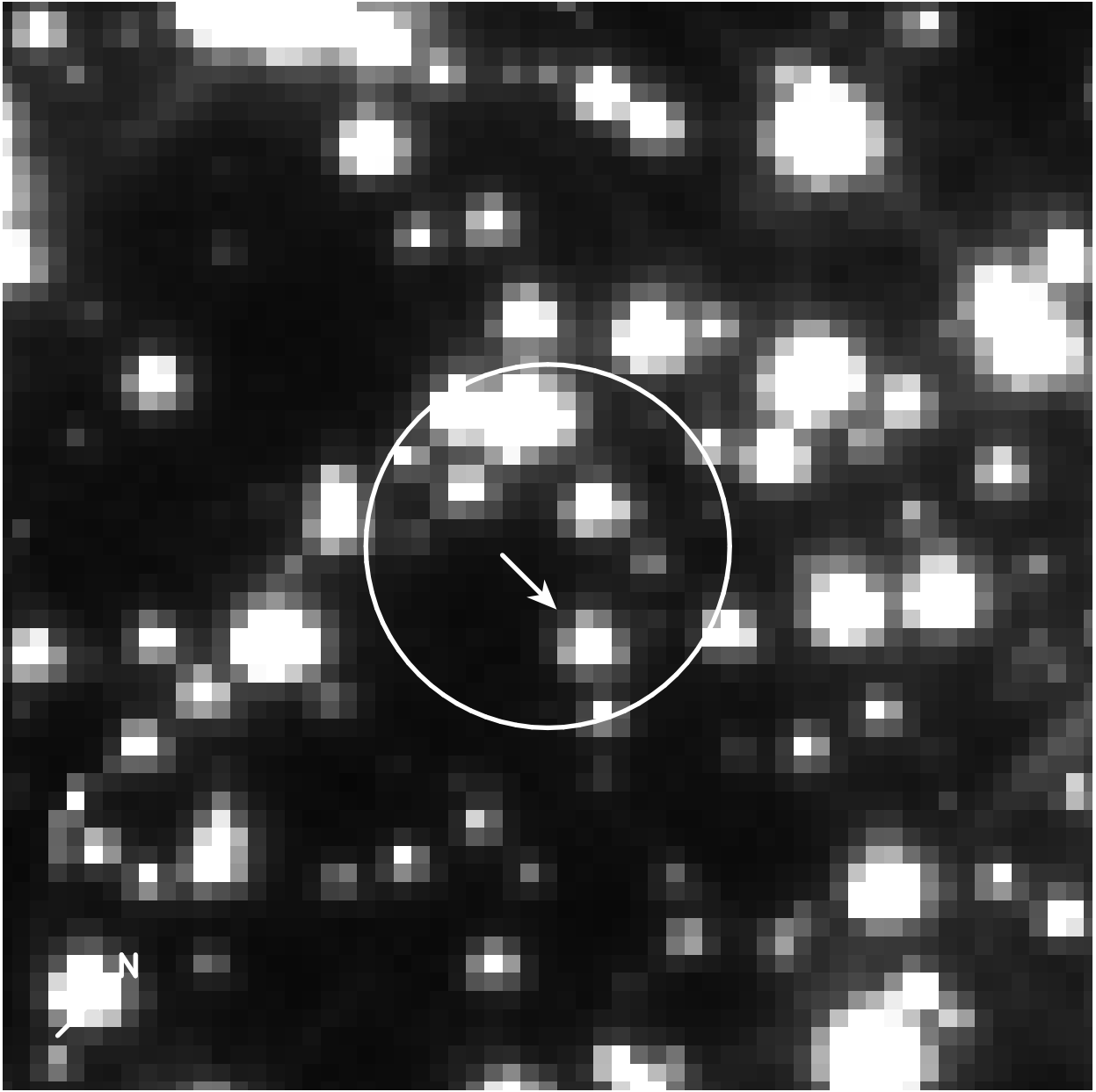}
\caption{Error circle of the position measurement of the event 
shown in Fig.~\ref{fig3} superimposed on a $6\arcsec\times6\arcsec$ section
of an HST image of the field (archival image u5330405r in F606W). The star indicated by the arrow, 
is the source star studied in detail by \citet{sahu01} and \citet{anderson03}. \label{fig4}}
\end{figure}

It is clear that the lightcurve shown in Fig.~\ref{fig3} is not microlensing, and is 
characteristic of multiple outbursts of a cataclysmic variable. We now examine whether or not
this is the same event reported by \citet{sahu01}. The object is in one of the standard MOA survey fields
for which position calibration information has been obtained \citep{bond01}. Our measurement of the centroid
position of the object then corresponds to a position in the sky of
$\alpha=18~36~24.70$ and $\delta=-23~54~35.1$ with an associated error circle of 
radius $\sim$1\arcsec. The uncertainty corresponds to the one standard deviation width of the model
Gaussian profile. In Fig~\ref{fig4}, we have
superimposed this error circle on an image from the HST program that observed M22 in 1999. The
resolved star that underwent the brightening episode observed by \citet{sahu01} is well within 
the error circle. It is conceivable, that the source star for the outburst events shown in 
Fig.~\ref{fig3} was some other star within the error circle. However, we consider this very
unlikely given that the same star identified by \citet{sahu01} was also identified as a
candidate cataclysmic variable by \citet{anderson03}. We therefore conclude that the observed
outbursts presented here and the episode reported by \citet{sahu01} correspond to the same
source star that was studied in detail by \citet{anderson03}, and that this star is a
dwarf nova system.

\section{Discussion}

Our observations, combined with those of \citet{sahu01}, exhibit a dwarf nova system that underwent at least
three outburst episodes during 1999--2004. This is the highest number of outbursts observed for a CV in
a globular cluster. The only other cases of multiple outbursts in such systems was that of V2 in 47 Tuc 
where two outbursts separated by 18 months were seen in a set of HST observations \citep{paresce94, shara96},
and an eruption during June 1985 lasting for more than 10 days of V101 in M5 \citep{shara87} followed by 
another outburst observed in June 1988 \citep{naylor89}.

%% THIS NEEDS rewriting
Our observations, together with those of \citet{sahu01} show that this CV in M22 has outburst episodes of varying
amplitudes. The first episode, reported by \citet{sahu01} attained a peak magnitude of $I\approx15$. 
In the second outburst in 2002, reported here,
only one sampling point was possible and so that measurement corresponds to a lower limit on the
peak luminosity of $I\le15$. This episode was therefore at least as bright as the first.
The third outburst
episode, also reported here, was of relatively low amplitude peaking at $I\approx15.7$. 
Here, we measured a FWHM of 14 days similar that of the episode reported by \citet{sahu01}.
The quiescent state magnitude of this object obtained by \citet{sahu01} from HST observations 
is $I_{\rm F814}=17.85$. The outburst amplitudes were then infered by subtracting this value from the calibrated
magnitudes of the peak luminosities. In all cases
the outburst amplitude in the range $\Delta I=$2--3 are of the order expected for dwarf novae of stars
in the field. Given the quiescent state absolute magnitude of $M_{\rm I}=4.7$ \citep{anderson03}, this object
therefore attains absolute magnitudes brighter than $M_{\rm I}\sim1.7$ during outburst.
The outburst episodes appear semi-regular with a recurrence rate of once every 1--2 years---consistent 
with the frequency of "superoutbursts" among dwarf novae in the field \citep{warner87}. 

In observations using XMM-Newton, \citet{webb04} observed an absorption feature at around 1 keV in an X-ray source
with an associated 5\arcsec~radius error circle that encloses the position of the optical position
of this dwarf nova. If this is interpreted as cyclotron resonance, this would place this object in the "intermediate
polar' subclass of dwarf nova in which matter is accreted from a secondary onto a moderately magnetized 
compact object \citet{warner83}. Amongst dwarf novae the compact objects are commonly white dwarfs. However, there
exists the interesting possibility that this dwarf nova system harbours an accreting neutron star 
primary. \citet{webb04} note that the outburst profile observed by 
\citet{sahu01} resembles that of the optical light curve of some X-ray transients. However, counting
against this hypothesis is the fact that the outburst amplitudes of X-ray transients of $>7$ mag are
considerably higher than observed here and they recur on much longer timescales of several years.
Identifying the nature of the compact object in this system would be of significant interest as this would
impact on our understanding of the formation and evolution of binary systems in globular clusters.

In this letter, we have confirmed that an event in M22 originally thought to be microlensing and later
suspected of being a cataclysmic variable, is in
fact a cataclysmic variable. Having also confirmed that outburst episodes repeat, further close monitoring
of this event for future outburst episodes is important. Time resolved X-ray and optical spectroscopy of
the source during both quiescent and outburst states are encouraged. 

%% Include an acknowledgments section in your paper,

\acknowledgments

The MOA project is supported by the Marsden Fund of New Zealand, the 
Ministry of Education, Culture, Sports, Science, and Technology (MEXT) 
of Japan, and the Japan Society for the Promotion of Science (JSPS).

\clearpage

\clearpage

%% Here we use \plottwo to present two versions of the same figure,
%% one in black and white for print the other in RGB color
%% for online presentation. Note that the caption indicates
%% that a color version of the figure will be available online.
%%

%% This figure uses \includegraphics to scale and rotate the still frame

\clearpage

\begin{table}
\begin{center}
\caption{I band photmetry during the two outburst episodes shown in Fig.~\ref{fig3}.\label{tab1}}
\begin{tabular}{cc}
\tableline\tableline
JD & I\\
\tableline
2570.933062 & 15.12$\pm$0.02\\
2861.868189 & 16.66$\pm$0.14\\
2861.940330 & 16.48$\pm$0.24\\
2868.941626 & 15.94$\pm$0.07\\
2869.008351 & 15.70$\pm$0.07\\
2869.861221 & 15.75$\pm$0.05\\
2869.933721 & 15.84$\pm$0.09\\
2874.862240 & 16.48$\pm$0.12\\
2875.904480 & 16.30$\pm$0.12\\
2876.028102 & 16.16$\pm$0.15\\
\tableline
\end{tabular}
\end{center}
\end{table}

\end{document}